# A FORMAL THEORY ON PROBLEM SPACE AS A SEMANTIC WORLD MODEL IN SYSTEMS ENGINEERING

Mayuranath SureshKumar[1] and Hanumanthrao Kannan[1]
[1]The University of Alabama in Huntsville

## ABSTRACT

*Classic problem-space theory models problem solving as a navigation through a structured space of states, operators, goals, and constraints. Systems Engineering (SE) employs analogous constructs (functional analysis, operational analysis, scenarios, trade studies), yet still lacks a rigorous systems-theoretic representation of the problem space itself. In current practice, reasoning often proceeds directly from stakeholder goals to prescriptive artifacts. This makes foundational assumptions about the operational environment, admissible interactions, and contextual conditions implicit or prematurely embedded in architectures or requirements. This paper addresses that gap by formalizing the problem space as an explicit semantic world model containing theoretical constructs that are defined prior to requirements and solution commitments. These constructs along with the developed axioms, theorems and corollary establish a rigorous criterion for unambiguous boundary semantics, context-dependent interaction traceability to successful stakeholder goal satisfaction, and sufficiency of problem-space specification over which disciplined reasoning can occur independent of solution design. It offers a clear distinction between what is true of the problem domain and what is chosen as a solution. The paper concludes by discussing the significance of the theory on practitioners and provides a dialogue-based hypothetical case study between a stakeholder and an engineer, demonstrating how the theory guides problem framing before designing any prescriptive artifacts.*

## I. INTRODUCTION AND BACKGROUND

Classic problem space theory [1] posits that people solve problems by mentally navigating a "problem space," a cognitive map containing the initial state, goal state, and all possible intermediate states and operators (actions) that transform one state to another, using heuristics (mental shortcuts) to search for the solution path, like finding your way through a maze. Systems Engineering (SE) has long adopted analogous ideas through constructs such as functional analysis, operational analysis, scenarios, and trade studies [3-8]. However, despite this conceptual alignment, SE lacks a rigorous systems-theoretic representation of the problem space itself [9, 10]. Agencies including NSF, NASA, and DARPA, have acknowledged these concerns in workshops focused on challenges and opportunities in systems engineering and design [11-15]. These efforts consistently point to a critical concern in problem space representation and reasoning highlighting the need for theoretical foundations needed to move SE beyond its traditional practice-driven state [16-18].

In current SE practice, reasoning about the problem frequently proceeds directly from stakeholder goals toward requirements and architectural concepts [19, 20, 21]. In doing so, fundamental assumptions about the operational environment, system boundaries, admissible interactions, and contextual conditions are left implicit or embedded prematurely within prescriptive artifacts [22-24]. As a result, early engineering decisions are often justified relative to unstated interpretations of the operational world rather than a shared and formally analyzable representation of the problem domain [25].

This absence of an explicit problem-space representation gives rise to several conceptual gaps [16, 17]. First, system boundaries are often treated as informal modeling conveniences rather than as semantic commitments that determine what is considered internal, external, or environmental, and which interactions must be accounted for when reasoning about system behavior [26]. Second, interactions are typically represented structurally, without a clear distinction between interactions that are merely possible and those that become active under specific operational conditions [27]. Third, properties of the domain, such as environmental constraints, admissible phenomena, and contextual dependencies, are frequently conflated with solution choices, becoming embedded within requirements or architectures rather than represented as properties of the problem space itself [28].

A central contributor to these gaps is the lack of a clear distinction between *what is true of the problem domain* and *what is chosen as a solution* [16, 17]. Domain truths include the existence of external systems, the structure of the operational environment, the admissibility of interactions across boundaries, and the conditions under which those interactions may occur. These properties exist independently of how a system is ultimately realized. Solution choices, in contrast, concern how the system-to-be is structured or implemented [29]. When this distinction is not made explicit, assumptions about the world are inadvertently treated as design decisions, and design decisions are later defended as if they were inherent properties of the domain. This circularity undermines disciplined reasoning about feasibility, correctness, and stakeholder intent [27].

Reasoning about the problem space also requires an explicit account of *what follows from interactions under context [27, 29, 30]*. When a system interacts with external systems and its environment under particular operational conditions, certain observable conditions arise as outcomes of those interactions. These outcomes, whether they correspond to desirable or undesirable effects cannot be meaningfully discussed without first specifying the boundary, the interacting entities, and the operational context under which those interactions occur. Without this structure, claims about the consequences of interactions under specific operational conditions remain informal and interpretation-dependent [27].

These gaps cannot be resolved by refining requirements or architectures alone [31]. Requirements and solutions are prescriptive artifacts that assume an underlying understanding of the operational world in which they are to function [3, 5, 19]. When that understanding is implicit or inconsistent, prescriptive artifacts inherit those deficiencies. As a result, validation and correction are deferred until late in the lifecycle, after significant design commitments have been made [32].

Addressing these limitations requires a formal representation of the problem space itself, one that makes domain assumptions explicit and supports rigorous reasoning about boundaries, interactions, and context prior to solution design [33].

## 2. Problem Space as a World Model

This paper therefore treats the problem space as a *world model*: a formally defined domain in which entities, boundaries, interactions, operational contexts, and their consequences are explicitly represented. In this view, the problem space is not defined by goals or solutions alone, but by the structured conditions under which systems and their environments interact. Outcomes, i.e., conditions that support or undermine stakeholder goals, are understood as consequences of these interactions under specific contexts, rather than as intrinsic properties of a system in isolation.

By making boundaries, interactions, and operational contexts explicit, the problem space becomes a domain over which rigorous reasoning is possible before requirements are specified or solutions are proposed. The objective of this work is not to prescribe solutions, define requirements, or compare alternative implementations. Instead, it establishes a foundational, systems-theoretic semantics for representing and reasoning about the problem space itself.

**Research Questions**

Grounded in the above motivation, this paper addresses the following research questions:

**RQ1:** *How can system boundaries, entities, and interactions be formally defined as semantic commitments that unambiguously constrain what is considered internal, external, and admissible in the problem space?*

This research question addresses a foundational ambiguity in current systems engineering practice: system boundaries are often treated as informal modeling conveniences rather than explicit semantic commitments. As a result, it is frequently unclear which entities and interactions must be accounted for when reasoning about the problem domain, leading to hidden assumptions and inconsistent interpretations across analyses. By asking how boundaries, entities, and interactions can be formally defined as semantic commitments, this work seeks to establish a precise and unambiguous basis for distinguishing internal, external, and environmental elements, as well as the admissibility of interactions among them. Resolving this question is essential for ensuring that all subsequent reasoning about context, outcomes, and sufficiency is grounded in a shared and explicit representation of the problem space, rather than in analyst-dependent interpretations.

**RQ2:** *How do operational contexts determine which interactions become active and how outcomes are grounded in explicit interaction sets rather than implicit or solution-specific assumptions?*

This research question targets a common source of confusion in early systems engineering reasoning: the implicit assumption that structurally defined interactions necessarily occur, and that outcomes follow directly from architecture or intent. In practice, interactions that are possible in principle may or may not be realized depending on environmental conditions, and outcomes are often justified through narrative explanations rather than explicit problem-space structure. By focusing on the role of operational context, this work seeks to formalize how environmental inputs select active interactions from the set of structurally available ones, and how outcomes can be grounded in explicit sets of such interactions. Addressing this question enables disciplined reasoning about what actually occurs under specific conditions, ensuring that outcome claims are traceable to the modeled problem space rather than to solution-specific or informal assumptions.

**RQ3:** *What does it mean for a problem-space representation to be sufficient for reasoning about desired outcomes across operational contexts, and how does this sufficiency evolve as outcomes, contexts, or system boundaries change?*

This research question addresses the challenge of reasoning rigorously about a problem space that is inherently incomplete and subject to evolution. In early phases of system development, stakeholders' desired outcomes, relevant operational contexts, and even the system-of-interest itself may change over time. Yet engineers still require criteria for determining when the problem-space representation is sufficiently specified to support meaningful reasoning. By framing sufficiency as an outcome-relative and context-dependent property, this work seeks to characterize what must be represented in the problem space to determine desired outcomes without assuming a definitive or final formulation. Addressing this question clarifies how sufficiency can be established, how it may be invalidated by new stakeholder concerns, and how consistent reasoning can be preserved across boundary re-selection and levels of decomposition.

Together, these research questions define the scope of this paper as the development of a formal foundation for representing and reasoning about the problem space in Systems Engineering, independent of requirements specification and solution design. Rather than focusing on prescriptive artifacts such as requirements, the questions collectively address how boundaries, interactions, operational contexts, and outcomes must be represented to support disciplined, context-aware reasoning about the operational world itself. By resolving these questions, the paper establishes the problem space as an explicit semantic domain over which consequences can be evaluated and sufficiency can be assessed prior to architectural or requirements commitments.

This paper is organized as follows: Section II presents the theory-development methodology, including the rationale for the selected formal foundations (systems theory, set theory, and propositional logic) and the process by which axioms, definitions, and theorems are iteratively developed. Section III establishes the formal problem-space domain constructs, and their relationship to stakeholder goals, enabling traceable attribution of consequences to explicit problem-space structure rather than implicit assumptions. Section IV consolidates key theoretical results and discusses their significance from a practitioner's perspective. Section V provides a hypothetical case study that applies these theoretical contribution as foundational semantics for problem space representation and reasoning in systems engineering. Section VI concludes the paper with future scope and present limitations.

## II. METHODOLOGY

The formal theory given in this paper is developed through a unique methodology that interlinks structured formalism with heuristic intuition. Unlike traditional approaches where a methodology precedes theory, this process was built iteratively through the process of theorization itself, driven by both practical insight and academic discourse. At its core, the methodology draws inspiration from Wacker's methodology for systematic theory development [34], the Axiomatic method popularized by David Hilbert 35] and the V-model commonly used in systems engineering [36]. These inspirations are adapted to the context of formal theoretical development in systems engineering practice to propose the methodological framework underpinning the development of our theory as shown in Figure 1.

The process begins with the identification of a real-world problem or need. This "problem definition" stage provides the motivational context and background for the theory development. Following this, formal languages must be selected based on which the theory is expressed and structured. A critical consideration here is whether the theory remains within the bounds of natural language (e.g., English) or employs more rigorous formalisms. Formalism addresses ambiguities, traceability issues, and incomplete decomposition through clear definitions, traceable mappings, and verifiable proofs [37, 38]. By augmenting systems theory with set theory and propositional logic, practitioners can rigorously assess equivalence between functions, ensuring successful function substitution. The reasoning for these selections is as follow:

*Systems Theory Augmented with Set Theory:* Systems theory provides a way to understand functions as parts of a larger whole, rather than in isolation [23, 39, 40, 41]. Wymore's systems theory provides a mathematical foundation for system design [23], Mesarovic's systems theory focuses on the hierarchical and multilevel organization of complex systems [39], Bertalanffy's general systems theory emphasizes the holistic nature of systems [40], and DEVS offers a formal framework for modeling discrete event systems [41]. Across these systems-theoretic frameworks, one principle remains consistent: a System Solution cannot be meaningfully understood in isolation, but only through its boundary-crossing interactions, operational contexts, and the outcomes it realizes relative to stakeholder goals. Understanding these system-level relationships is essential for reasoning about the problem space world. When we combine general systems theory with set theory, we gain precise ways to describe these relationships. Systems theory by itself is not formally stated [39, 40] but set theory provides the necessary formal constructs to represent and reason about concepts proposed by systems theory.

*Propositional Logic:* Formal logic was developed in the late 19th and early 20th centuries to model reasoning with mathematically precise structures [42]. Modern formal logics are comprised of three components: 1) a set of recursively defined sentences or symbolic representations for a set of base symbols that make up a formal language; 2) a precise and rigid semantics that gives meaning to the sentences; and 3) a proof theory that connects a set of sentences (premises) to another sentence (conclusion) [43]. Within their respective domains, formal logic serves as a powerful tool for reasoning. Propositional logic for example, can be applied to domains ranging from those involving only simple propositions to scenarios that involve modalities like necessity, always will be, prefers, sufficiency, knows, believes, etc. We leverage propositional logic as a proof mechanism to derive theorems and corollaries from the formal definitions established using systems theory, augmented by set theory, in the following section.

Following this, based on the formal language and our need, a set of axioms is formulated. These axioms represent the foundational assumptions upon which the rest of the theory is built. These axioms are not empirically tested but are accepted as foundational truths within the context of the problem being addressed. It is important to recognize that intuition and heuristic observation often precede formalism. Questions such as, *"We observe a pattern, can we define it formally using logic?"* or *"What additional structure is needed to formally explain why the given solution emerges as it does?"* reflect the natural progression from informal insight to formal theory development. Such statements highlight the symbiotic relationship between observed patterns and theoretical support, suggesting that intuition and belief act as precursors to logical rigor in any developed theory.

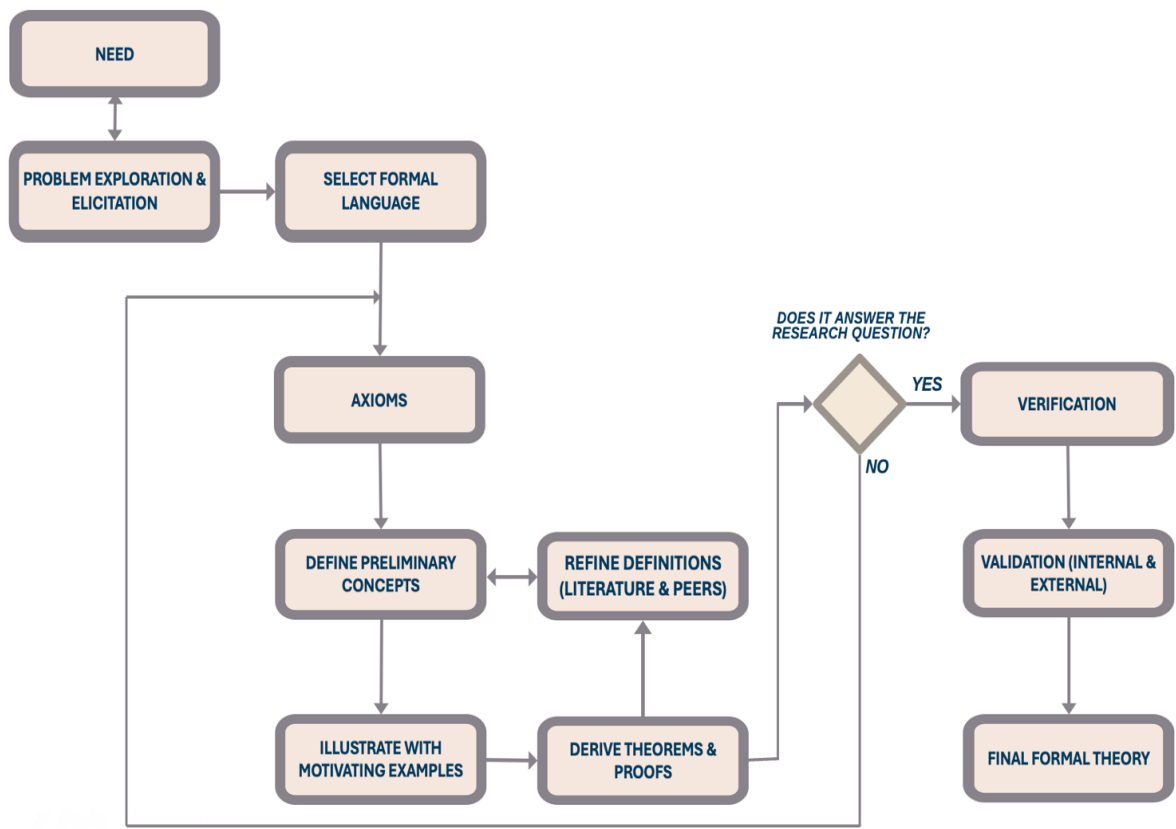

*Figure 1. Methodology on development of proposed theory*

The next phase involves defining preliminary definitions. These are often borrowed or adapted from existing literature to provide a common grounding. To ensure conceptual alignment, iterative peer discussions are conducted. This collaborative aspect is crucial for ensuring alignment of shared mental models in the developed theory. To aid in clarity and communication, illustrative examples are constructed alongside the definitions given in Section III. A critical component of this methodology is its iterative nature. Throughout the process, the research question serves as a central reference point. As the theory develops, if a derived theorem or its implications fail to address the core research question, the process returns to earlier stages: revisiting axioms, refining definitions, or redefining theorems and proofs. Finally, once the derived theorems are shown to consistently support all the research questions and ensure verification and validation through mathematical logical soundness, relevant supporting examples, discussions, and peer review, the theory is considered complete. This gives a formal theory that is both rigorously constructed and contextually grounded, ready for application.

### III. FORMAL PROBLEM SPACE CONSTRUCTS

To move from intuitive reasoning about problem space representation to a rigorous, repeatable theory, the concepts involved must be grounded in precise and unambiguous foundations. One cannot rely solely on informal descriptions or example-based intuition because the underlying assumptions must be stated explicitly so that all subsequent definitions and theorems follow logically and consistently. For this reason, the development of the theory begins by establishing a small set of fundamental axioms that formalize the essential truths on which the rest of the framework is built.

*Axiom 1 (Closed World Sufficiency):*
*All formal constructs that define the problem space world exist within the closed-world boundary, and any phenomenon irrelevant to problem-space semantics and reasoning is assumed outside the boundary.*
*Note:* The closed-world boundary defines the scope of representation of the problem-space world and is distinct from the system boundary B (Definition 9), which partitions the modeled world into the system solution SysSol and external entities $ES \cup Env$.

*Axiom 2 (Boundary Commitment):*
*Any entity, subsystem, or composite system within the closed-world boundary may be designated as the system of interest by explicitly defining a system boundary that partitions internal elements from external entities.*
*Note:* This boundary assignment must preserve the semantics of interactions, such that admissible inputs, outputs, and interaction directionality remain well-defined with respect to the newly designated system of interest.

*Axiom 3 (Outcome Truth):*
*Any outcome defined within the problem-space world has a truth value only with respect to a specified operational context.*

The rationale for these axioms is discussed in Section V, where their role in enabling disciplined problem space reasoning is made explicit. We begin with these axioms, because they capture the irreducible assumptions needed to reason about problem space representation in a formal framework. All core notions are introduced later as definitions and theorems, to avoid embedding unnecessary assumptions at the axiomatic level. This choice ensures minimality, i.e., the axioms do no more than establish ground truths, and all richer concepts are built on top of them.

The above axioms have laid the foundation for the rest of this section. The next step is to formalize the fundamental concepts that constitute the problem-space world model. These theoretical constructs will serve as the basis for reasoning about the problem space world. They enable disciplined reasoning about what can occur and what follows under specified conditions before solution architectures are proposed. *Definitions 1-17* are formally developed using the above-mentioned systems theory concepts [23, 39, 40] to ensure precision and logical consistency. Within the problem space world model, internal system functions represent what the system does by transforming admissible inputs into measurable outputs [44]. Since these inputs and outputs are the boundary-crossing primitives from which interactions, contexts, and outcomes can be inferred, we begin by formalizing what inputs and outputs mean.

*Definition 1 (Inputs):*
*Inputs, denoted by I, are the admissible signals or exchange of material or energy that cross the boundary (Definition 9) or are generated within the boundary as outputs of preceding functions (Definition 16) that serve as the inputs to subsequent functions.*
*Conditions:*
- *For every admissible input $i$, there exists at least one function $f \in F$ such that $i \in Dom(f)$. This ensures that all admissible inputs have a well-defined functional transformation.*

*Note:* In this paper, '*admissible*' means a flow is well-formed and permitted by the problem-space semantics, i.e., it can legally propagate through a defined set of interactions (Definition 3) and lies within the receiving entity's designed input domain under the specified boundary (Definition 9) and operational context (Definition 11).

*Definition 2 (Outputs):*
*Outputs, denoted by O, are measurable input transformations produced from signals or exchanges of material or energy by executing at least one function ($f \in F$, Definition 16).*
*Conditions:*
- *If an output $o$ remains inside the boundary, then it may serve as an input to another internal function $g \in F$ only if $o \in Dom(g)$.*
- *If an output $o$ crosses the boundary and enters an external system $E$, then it may serve as an input only to those functions in $E$ whose admissible input domains admit the output. Formally, $o \in Dom(f_E)$ where $f_E$ is the specific receiving function inside the external system.*

*Definition 3 (Interaction):*
*An interaction relation, denoted by IR, is a directed relation $IR \subseteq E \times E$, where $E = |SysSol| \cup ES \cup Env$. Each ordered pair $(x, y) \in IR$ denotes a potential exchange of a flow from source element $x$ to destination element $y$.*
A *flow* is a signal, material transfer, energy transfer, or environmental influence represented as a value that can be produced by a source element and interpreted by a receiving function.
Here, $|SysSol|$ denotes the underlying set of internal system functions contained in the System Solution. The full System Solution, introduced later in Definition 4, is a structured entity that includes both this underlying function set and its internal interaction relation $IR_{int}$.
*Conditions:*
- *If the destination element $y$ contains a receiving function $f_y$, then a flow may propagate along $(x, y)$ only if $o(x, y) \in Dom(f_y)$. This ensures that only flows the receiving function can interpret are permitted to propagate along the interaction.*
- *Each interaction $(x, y)$ may be realized through an interface that carries a flow value $o(x, y)$, provided the admissibility condition above is satisfied. The interface specifies the modality, type, or physical/logical channel through which the flow is exchanged, ensuring that the flow is well-formed.*

*Note:* The relation *IR* specifies the structural possibility of interactions. The Operational contexts (Definition 11) determine which interactions become active, meaning which interactions actually carry flows under a given input. Elements of *ES* and *Env* are treated as function-like black-box entities that may produce or receive flows through boundary-crossing interactions. Their internal structure is not modeled explicitly, and only the admissibility of flows to their receiving functions (where applicable) is considered. In contrast,

the System Solution $SysSol$ is modeled as a structured entity whose underlying carrier set is $|\,SysSol\,|$. The distinction allows $E$ to be defined cleanly while maintaining different abstraction levels for internal functions and external/environmental entities.

*Definition 3.a (Internal Interaction):*
*An internal interaction, denoted by $IR_{int}$, is an interaction whose source and destination both lie within the System Solution (SysSol). Formally, $(x, y) \in IR_{int}$ iff $x \in |SysSol|$, $y \in |SysSol|$. Each internal interaction is realized through an interface inside the system boundary that carries a flow value $o(x, y)$.*
*Conditions:*
*The admissibility and interface-binding conditions for internal interactions follow directly from Definition 3.*

*Definition 3.b (Boundary-crossing Interaction):*
*A boundary-crossing interaction is an interaction whose source and destination lie on opposite sides of the system boundary.*
**Inbound Boundary-Crossing Interaction:** *$(x, y) \in IR_{in}$ iff $x \in ES \cup Env$, $y \in |SysSol|$.*
**Outbound Boundary-Crossing Interaction:** *$(x, y) \in IR_{out}$ iff $x \in |SysSol|$, $y \in ES$.*
*Each boundary-crossing interaction is realized through an interface at the system boundary and carries a flow value $o(x, y)$.*
*Conditions:*
*The admissibility and interface-binding conditions for boundary-crossing interactions follow directly from Definition 3.*

*Definition 3.c (External Interaction):*
*An external interaction, denoted by $IR_{ext}$, is an interaction whose source and destination both lie outside the System Solution. Formally, $(x, y) \in IR_{ext}$ iff $x \in ES \cup Env$, $y \in ES \cup Env$. Each external interaction is realized through an interface external to the system boundary and carries a flow value $o(x, y)$.*
*Conditions:*
*The admissibility and interface-binding conditions for external interactions follow directly from Definition 3.*

*Definition 4 (System solution):*
*A System Solution, denoted SysSol, is a structured entity consisting of:*
- *a set of internal system functions $F$, and*
- *an internal interaction relation $IR_{int} \subseteq F \times F$.*

*Formally, $SysSol = (F, IR_{int})$, and its underlying carrier set is defined as $|\,SysSol\,| = F$.*
*Conditions:*
- *Any flow produced by a function in $F$ that does not cross the system boundary must be admissible to another function in $F$. This ensures all internal flows remain well-formed within the System Solution.*
- *Any flow that leaves the System Solution must do so through an outbound boundary-crossing interaction in $IR_{out}$. This ensures externalized flows use defined boundary interfaces.*
- *Any flow entering the System Solution must arrive via an inbound boundary-crossing interaction in $IR_{in}$. This ensures all external influences enter through well-formed interfaces.*

*Properties:*
1. ***Open System:*** The System Solution is an open system; at least one admissible input or output crosses the system boundary. This ensures meaningful interaction with external systems or the environment.

2. *Functional Abstraction:* The System Solution may be abstracted as a high-level function SysSol: $I_b \rightarrow O_b$, where $I_b$ is the set of inbound flows and $O_b$ is the set of outbound flows.

*Note:* $|SysSol|$ is used in Definition 3 to denote the carrier set of internal system functions. The boundary-crossing relations $IR_{in}$ and $IR_{out}$ are defined in Definition 3.b and are not part of the internal structure of $SysSol$ but represent interactions across the system boundary.

*Definition 5 (External Systems):*
External systems, denoted ES, are system entities that lie outside the system boundary yet participate in boundary-crossing interactions with the System Solution. Formally, $ES \subseteq E \setminus |SysSol|$.
*Conditions:*
- For any outbound interaction $(x, y) \in IR_{out}$ with $x \in |SysSol|$ and $y \in ES$, the flow must satisfy $o(x, y) \in Dom(f_y)$ for some receiving function $f_y$ internal to the external system. This ensures that outbound flows are admissible to external receiving functions.
- For any inbound interaction $(x, y) \in IR_{in}$ with $x \in ES$ and $y \in |SysSol|$, the flow must satisfy $o(x, y) \in Dom(f_y)$. This ensures that inbound flows are admissible to internal functions of the System Solution.

*Note:* External systems are treated as black-box functional entities whose internal structure is not modeled. Their role is limited to producing or receiving flows through boundary-crossing interactions defined in Definition 3.b.

*Definition 6 (Operational Environment):*
The operational environment, denoted Env, is the set of exogenous entities that lie outside the System Solution and external systems but may affect them through inbound boundary-crossing interactions. Environmental influences appear as flows generated by elements of Env through interactions of the form: $(x, y) \in IR_{in}, x \in Env, y \in |SysSol| \cup ES$.
An environmental input is any flow produced by an environmental entity that enters the System Solution or an external system via such interactions.
*Conditions:*
- $Env \cap |SysSol| = \emptyset, Env \cap ES = \emptyset$. This ensures that environmental entities remain distinct from both the internal system and the external systems.
- Environmental factors may affect the System Solution or external systems only through inbound boundary-crossing interactions. This ensures that environmental effects never bypass the system boundary or its defined interfaces.

*Note:* The internal structure of environmental entities is not modeled. They are treated as black-box sources of exogenous flows. The environment itself is not an '*input*', rather, it produces inputs in the form of flows that enter through defined interactions.

*Definition 7 (States):*
A state, denoted $s$, of an entity $y$ (internal system function or external system) is an internal configuration that, together with admissible inputs, determines how that entity may evolve. For each $y \in |SysSol| \cup ES$, its state space $S_y$ is a non-empty set of all internal configurations it may occupy. At any given instant, the entity is in exactly one state $s_y \in S_y$.
*Conditions:*
- Every admissible internal configuration of $y$ must belong to its declared state space $S_y$.
- If state transitions (Definition 8) are defined for $y$, then for each state $s_y \in S_y$ and each admissible input to $y$, the resulting internal configuration must also lie within $S_y$.

*Note:* An *internal configuration* refers to internal variables, modes, or conditions of $y$ that influence how it responds to admissible inputs but are not themselves exchanged across interactions. This distinguishes states from flows, inputs, outputs, and outcomes, which may cross system boundaries. For example, in a thermostat controller, internal configurations such as '*heating*,' '*cooling*,' or '*idle*' determine how temperature inputs are interpreted and which outputs are produced, yet these internal modes are never transmitted externally; only the resulting control command is exchanged.

*Definition 8 (State Transitions):*
*A state transition for an entity $y$ is a function describing how its internal configuration changes when it receives an admissible input. Let $S_y$ be the state space of $y$, and let $I_y$ be the set of admissible inputs delivered to $y$ via interactions satisfying the admissibility condition of Definition 3. A state transition function for $y$ is a mapping: $\tau_y: S_y \times I_y \to S_y$.*
*Conditions:*
- *For all $s_y \in S_y$ and all $i \in I_y: \tau_y(s_y, i) \in S_y$. This ensures that transitions never yield undefined internal configurations.*
- *The transition function $\tau_y$ is evaluated only on inputs arising from flows $o(x, y)$ satisfying $o(x, y) \in Dom(f_y)$. This ensures that transitions occur only under well-formed interactions.*

*Theorem 1 (Environmental Flow May Induce State Changes)*
*Let $y \in |\,SysSol\,| \cup ES$ be any entity with state space $S_y$ and an associated state-transition function $\tau_y: S_y \times I_y \to S_y$. If an environmental entity $x \in Env$ participates in an inbound interaction $(x, y) \in IR_{in}$ and the resulting flow $o(x, y)$ is admissible to $y$, then the environmental input may drive $y$ from one state to another.*
**Proof:** By Definition 6, environmental inputs manifest as flows entering entities in $|\,SysSol\,| \cup ES$ only through inbound boundary-crossing interactions. Hence $(x, y) \in IR_{in}$. By Definition 3, flow propagation along $(x, y)$ is permitted only if $o(x, y) \in Dom(f_y)$, which holds by hypothesis. By Definition 8, admissible inputs delivered to $y$ form the set $I_y$. Since $o(x, y)$ is admissible, $o(x, y) \in I_y$. Let $s_y \in S_y$ be the current internal state of $y$. Applying the state-transition function to this state and the admissible input yields $s'_y = \tau_y(s_y, o(x, y))$. By Definition 8, the transition function $\tau_y$ maps any state and any admissible input to an element of $S_y$, therefore $s'_y \in S_y$. Thus, an admissible environmental flow may induce a state transition for $y$ (the resulting state $s'_y$ may or may not differ from $s_y$).

*Theorem 2 (Environmental Flow May Activate Specific Interactions)*
*Let $\eta$ be an environmental input generated by some $x \in Env$. If $(x, y) \in IR_{in}$ and the resulting flow $o(x, y)$ is admissible to $y$, then the interaction $(x, y)$ is active under the environmental input $\eta$.*
**Proof:** By Definition 6, environmental inputs manifest as flows entering entities in $|\,SysSol\,| \cup ES$ only through inbound boundary-crossing interactions. Hence $(x, y) \in IR_{in}$. By Definition 3, a structural interaction becomes active when a flow is carried along it subject to the admissibility condition $o(x, y) \in Dom(f_y)$. Since admissibility holds, Definition 4 permits the interface associated with $(x, y)$ to carry the flow value $o(x, y)$. Under the environmental input $\eta$, this flow is in fact propagated along $(x, y)$ through that interface. By Definition 3, an interaction is active when a flow satisfying the admissibility condition is propagated along its interface. Because the environmental input $\eta$ produces the admissible flow $o(x, y)$ that is propagated along $(x, y)$, the interaction $(x, y)$ is realized under $\eta$. Therefore, the admissible environment flow activates the interaction $(x, y)$.

*Definition 9 (Boundary):*
*A boundary, denoted B, is a formally specified construct that partitions a universe of interacting entities $U$ into two disjoint sets: $B = (U_{int}, U_{ext}), U_{int} \cap U_{ext} = \emptyset, U_{int} \cup U_{ext} = U$. The boundary identifies which entities are treated as internal to the system of interest and which are treated as external.*

*Conditions:*
- *Relative to the boundary B, a signal, material, or energy flow is exchanged through either an internal interaction $IR_{int}$, a boundary-crossing interaction $IR_{in}$ or $IR_{out}$, or an external interaction $IR_{ext}$.*

Based on where the partition is created, the following boundary classification can exist.

### Definition 9.a (System Boundary):
*A system boundary, denoted $B_S$, is the boundary that partitions the System Solution from the External Systems and the Operational Environment: $B_S = (\;|SysSol|\;,\;ES \cup Env\;)$. Relative to the system boundary $B_S$, interactions are classified according to Definitions 3.a-3.c as internal, inbound, outbound, or external as shown in Figure 2.*
*Conditions:*
- *Any admissible flow exchanged between two entities inside the system boundary must occur through internal interactions, $(x,y) \in IR_{int} \mid x,y \in |SysSol|$*
- *Any admissible flow crossing the boundary $B_S$ is exchanged through boundary crossing interactions,*
    - ***Inbound if:*** *$(x,y) \in IR_{in}$ iff $x \in ES \cup Env$, $y \in |SysSol|$,*
    - ***Outbound if:*** *$(x,y) \in IR_{out}$ iff $x \in |SysSol|$, $y \in ES$*
- *Any admissible signal, material, or energy flow outside the boundary $B_s$ is exchanged through external interactions, $IR_{ext} = \{(e_1, e_2) \in IR \mid e_1, e_2 \in ES \cup Env\}$*

Relative to $B_S$, these interaction classes are mutually exclusive and collectively exhaustive. These conditions give the system boundary formal semantic meaning by constraining how signals, material, or energy may propagate across it. In all cases, flow propagation is subject to admissibility constraints as defined in Definition 3.

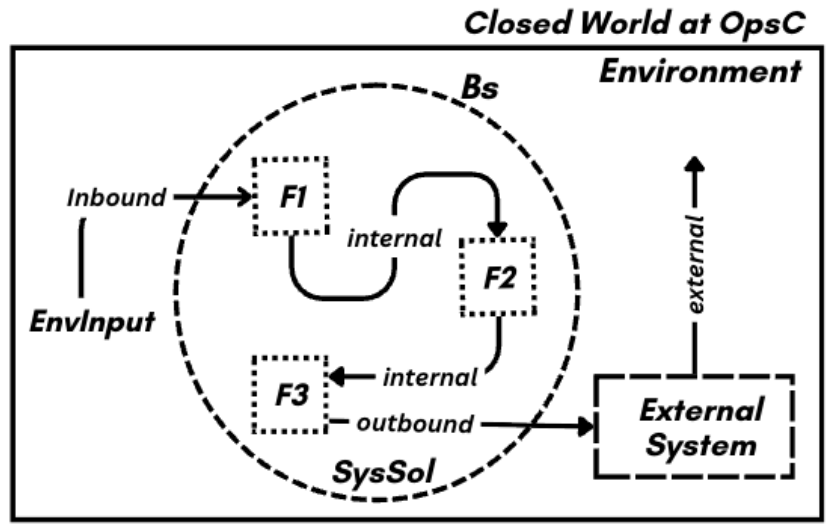

*Figure 2: Active Interactions in given Operational Context (OpsC)*

### Theorem 3 (Boundary Commitment May Recursively Be Applied)
*Any internal element or subsystem $z \in |SysSol|$ may be designated as a new system-of-interest by declaring a boundary around it. This induces the same four-class interaction classification (Definition 9.a) relative to the new boundary, and admissibility constraints (Definition 3) are preserved.*

**Proof:** By Axiom 2 (Boundary Commitment), any modeled entity within the closed-world boundary may be designated as a system-of-interest by explicitly defining a boundary that partitions internal from external elements while preserving interaction semantics. Let $z \in |SysSol|$ be selected as the new system-of-interest, and let $B_z$ be the boundary partitioning $Z$'s internal elements from all other entities. By Definition 9, $B_z$ is a valid boundary. By Definition 9.a, every interaction involving $Z$ is classified as internal, inbound, outbound, or external relative to $B_z$, and these classes are mutually exclusive and collectively exhaustive. By Definition 3, admissibility of a flow $o(x, y)$ depends only on whether $o(x, y) \in Dom(f_y)$. This condition is intrinsic to the receiving function and does not depend on boundary choice. Therefore, the same boundary semantics and admissibility constraints apply at the subsystem level.

Figure 3 illustrates how the system boundary may be treated as recursive. Once a $SysSol$ boundary $B_S$ is established, any internal element (e.g., the subsystem abstracted by $F_2$) may itself be treated as a new system-of-interest by placing a boundary $B_2$ around it, thereby re-partitioning what is inside vs. outside. This re-classifies the active interactions as inbound, internal, or outbound relative to $B_2$.

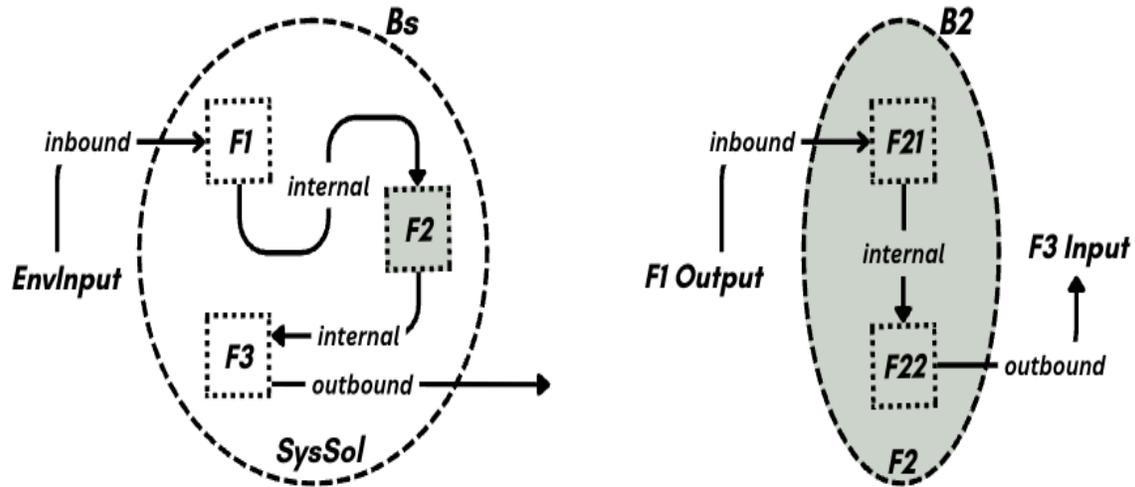

*Figure 3: Recursive Boundary Commitment (Axiom 2, Theorem 3)*

### Definition 10 (Operational solution):
*An Operational Solution, denoted $OpsSol$, is a structured entity that captures the configuration in which the System Solution interacts with External Systems and the Operational Environment through all admissible interaction relations. Formally, $OpsSol = (|\,SysSol\,| \cup ES \cup Env, IR_{int} \cup IR_{in} \cup IR_{out})$, where*
- *$|SysSol|$ is the set of internal system functions,*
- *ES is the set of external systems,*
- *Env is the operational environment,*
- *$IR_{int}$ contains internal interactions,*
- *$IR_{in}$ contains inbound boundary-crossing interactions, and*
- *$IR_{out}$ contains outbound boundary-crossing interactions.*

### Definition 11 (Operational Context):
*An Operational Context, denoted $OpsC$, is a realization of the Operational Solution under a specific environmental input. It identifies which structurally available interactions in the Operational Solution become active, i.e., actually carry flows, under that environmental condition. Formally, $OpsC = (EnvInput, IR^*)$, where*
- *EnvInput is a flow or collection of flows produced by one or more entities in Env through inbound boundary-crossing interactions, and*
- *$IR^* \subseteq IR_{int} \cup IR_{in} \cup IR_{out}$ is the set of interactions that are active under that environmental input.*

**Conditions:**

1. $IR^* \subseteq IR_{int} \cup IR_{in} \cup IR_{out}$. This ensures that an Operational Context selects only from the interactions structurally available in the Operational Solution.
2. For every active interaction $(x,y) \in IR^*$, $o(x,y) \in Dom(f_y)$. Only admissible flows may be propagated along active interactions.
3. If an environmental entity produces an admissible flow entering $y$, then the corresponding inbound interaction is active: $x \in Env$, $o(x,y) \in Dom(f_y) \Rightarrow (x,y) \in IR^*$. Environmental flows activate the interactions through which they propagate.

**Note:** Different environmental inputs may produce different operational contexts, even though the underlying Operational Solution remains fixed. Operational contexts govern which interactions are active at runtime; the System Solution and Operational Solution specify only the structural possibilities.

### Theorem 4 (Active Interactions Are Realized By Operational Context)
*The existence of an interaction in the problem space world does not imply that the interaction is active. Specifically, for any $(x,y) \in IR$, it need not be the case that $(x,y) \in IR^*$. An interaction is active only relative to a specified operational context.*

**Proof:** By Definition 11 (Operational Context), $IR^* \subseteq IR_{int} \cup IR_{in} \cup IR_{out}$ is the set of interactions active under $EnvInput$. Since $IR^*$ is a subset selected under context, interactions present in $IR$ does not guarantee presence in $IR^*$. Therefore, $(x,y) \in IR \nRightarrow (x,y) \in IR^*$.

### Definition 12 (Outcome):
*An Outcome (system outcome), denoted $o_c$, is a proposition describing a condition that arises within an Operational Context as a consequence of active interactions involving the System Solution.*
*Let,*
$$OpsC = (EnvInput, IR^*)$$
*be an Operational Context (Definition 11), and let $OC$ denote the set of outcome propositions represented in the system model. Each outcome $o_c \in OC$ is evaluated relative to an Operational Context and has an associated truth value:*
$$o_c: OpsC \to \{TRUE, FALSE\}.$$

### Conditions:
- *The truth value of each outcome $o_c$ is determined by the Operational Context $OpsC$ in which it is evaluated.*
- *Each outcome $o_c \in OC$ is associated with a set of active interactions $\mathfrak{I}_{o_c} \subseteq IR^*$ whose combined occurrence under $OC$ is sufficient to determine the condition described by $o_c$.*
- *Each outcome must involve participation of the System Solution: $\mathfrak{I}_{o_c} \cap \{(x,y) \in IR^* \mid x \in |SysSol| \lor y \in |SysSol|\} \neq \emptyset$. This ensures that outcomes are attributable to the System Solution and do not arise solely from interactions among external systems or environmental entities.*
- *The set $OC$ is specified as part of the system model and represents outcome propositions relevant to the system's purpose.*

*Note (Semantic Interpretation):* The mapping from Operational Contexts to outcome truth values constitutes a semantic interpretation of system operation. In this framework, semantics are introduced by interpreting active interactions and their induced effects at the level of the Operational Context, rather than by interpreting internal states or flows directly. This interpretation is constrained by system structure, boundary semantics, and interaction activation, ensuring that outcome evaluation is deterministic, traceable, and repeatable for a given Operational Context.

### Definition 12.a (Internal Outcomes):
*An outcome $o_c \in OC$ is an \*internal outcome\* if all interactions in its associated interaction set occur entirely within the System Solution boundary. Formally: $\mathfrak{I}_{oc} \subseteq IR^*_{int}$ where $IR^*_{int} = \{(x,y) \in IR^* \mid x \in |SysSol| \land y \in |SysSol|\}$. Internal outcomes describe conditions arising from interactions among internal system functions. These conditions are not directly observable from outside the system boundary.*

*Definition 12.b (External Outcomes):*
*An outcome $o_c \in OC$ is an \*external outcome\* if at least one interaction in its associated interaction set crosses the system boundary. Formally: $\mathcal{I}_{oc} \cap (IR_{in}^* \cup IR_{out}^*) \neq \emptyset$ where:*
- $IR_{in}^* = \{(x,y) \in IR^* \mid x \in ES \cup Env, y \in |\,SysSol\,|\}$
- $IR_{out}^* = \{(x,y) \in IR^* \mid x \in |\,SysSol\,|, y \in ES \cup Env\}$

*External outcomes describe conditions that manifest at or across the system boundary, arising from interactions between the System Solution and external systems or the environment. These conditions are observable from outside the system boundary.*

*Theorem 5 (Only Active Interactions May Affect the Problem Space World)*
Let $OpsC = (EnvInput, IR^*)$ be an operational context (Definition 11). Let $x, y \in E$. If $x$ affects $y$ in $OpsC$, then there exists an interaction $(x,y) \in IR^*$ and a flow value $o(x,y)$ such that $o(x,y) \in Dom(f_y)$. In particular, $x$'s affect on $y$ is realized only through an active interaction that carries an admissible flow into $y$.
*Note:* In this paper, '$x$ affects $y$' is grounded in at least one of the following meanings:
- **State-based Affect:** $y$ undergoes a state transition due to an input delivered from $x$
- **Outcome-based Affect:** the truth of some outcome proposition $o_c$ depends on an interaction involving $(x,y)$ in $IR^*$

*Proof:*
- **Case 1 (State-based Affect):** If x affects y by inducing a state change in y, then by Definition 8, y's transition function $\tau_y$ is evaluated on some input $i \in I_y$. By Definition 3, inputs in $I_y$ arise only as flows carried on interactions into y, and such a flow may propagate only if it is admissible, i.e., $o(x,y) \in Dom(f_y)$. Since the transition occurs in $OpsC$, the interaction carrying that flow is active, hence $(x,y) \in IR^*$. Therefore, $(x,y) \in IR^*$ and $o(x,y) \in Dom(f_y)$.
- **Case 2 (Outcome-based Affect):** If $x$ affects $y$ through its contribution to an outcome $o_c$, then by Definition 12, the truth of $o_c(OC)$ is determined by a set of active interactions $\mathcal{I}_{oc} \subseteq IR^*$. Since the influence is attributed to $x$ acting on $y$, $(x,y) \in \mathcal{I}_{oc}$, hence $(x,y) \in IR^*$. By Definition 3, any flow carried on $(x,y)$ must satisfy admissibility at the destination, so $o(x,y) \in Dom(f_y)$. Therefore, $(x,y) \in IR^*$ and $o(x,y) \in Dom(f_y)$.

*Note (Sufficiency):* From a mathematical standpoint, the term sufficiency mirrors a core logical concept of implication and contrapositive in propositional logic [43, 45, 46]. Sufficiency, formalized as $P \rightarrow Q$ (if P is true, then Q must be true) guarantees that when lower-level artifacts are satisfied, enough evidence is provided to establish confidence in the satisfaction of higher-level artifacts, creating a chain supported by logic. This yields a traceable linkage between artifacts, providing a logically sound foundation for disciplined theorem reasoning and proof.

*Theorem 6 (Outcome Invariance Under Irrelevant Context Variation)*
Let $o_c \in OC$ be an outcome and let $OpsC_1 = (EnvInput_1, IR_1^*)$ and $OpsC_2 = (EnvInput_2, IR_2^*)$ be two operational contexts under consideration. If there exists an interaction set $\mathcal{I}_{oc} \subseteq IR_1^* \cap IR_2^*$ that is sufficient to determine $o_c$ under both $OpsC_1$ and $OpsC_2$, then $o_c$ has the same truth value under $OpsC_1$ and $OpsC_2$.
*Proof:* By Definition 12, the truth value of $o_c$ under an operational context is determined by the occurrence of an associated interaction set sufficient to establish the outcome condition. If the same interaction set $\mathcal{I}_{oc}$ is active in both $OpsC_1$ and $OpsC_2$, and is sufficient to determine $o_c$ in each context, then the conditions required to establish $o_c$ are satisfied identically in both cases. Variations in other active interactions outside $\mathcal{I}_{oc}$ are irrelevant to the determination of $o_c$, since $\mathcal{I}_{oc}$ alone suffices. Therefore, the truth value of $o_c$ is invariant between $OpsC_1$ and $OpsC_2$.

***Theorem 7 (Existence of Minimal Interaction Sets for Outcome Determination)***
Let $o_c \in OC$ be an outcome and let $OpsC = (EnvInput, IR^*)$ be an operational context under consideration. If $o_c$ can be determined under $OpsC$, then there exists at least one interaction set $\mathcal{I}_{o_c}^{min} \subseteq IR^*$ such that:
1. $\mathcal{I}_{o_c}^{min}$ is sufficient to determine the truth value of $o_c$ under $OpsC$, and
2. For any strict subset $J \subset \mathcal{I}_{o_c}^{min}$, $J$ is not sufficient to determine the truth value of $o_c$ under $OpsC$.

*That is, outcome determination admits at least one minimal interaction set under the given operational context.*
**Proof:** By Definition 12, if $o_c$ can be determined under $OpsC$, then there exists at least one associated interaction set $\mathcal{I}_{o_c} \subseteq IR^*$ whose occurrence under $OpsC$ is sufficient to determine $o_c$. Consider the collection of all subsets of $\mathcal{I}_{o_c}$ that are sufficient to determine $o_c$ under $OpsC$. This collection is non-empty since it contains $\mathcal{I}_{o_c}$ itself. Partially order this collection by set inclusion. By standard set-theoretic reasoning, there exists at least one minimal element under this ordering. Let such a minimal element be denoted $\mathcal{I}_{o_c}^{min}$. By construction, $\mathcal{I}_{o_c}^{min} \subseteq IR^*$ is sufficient to determine $o_c$, and no strict subset of $\mathcal{I}_{o_c}^{min}$ is sufficient to do so. This means, removing any interaction from $\mathcal{I}_{o_c}^{min}$ yields a set that is no longer sufficient. Hence a minimal interaction set for determining $o_c$ under $OpsC$ exists.

***Theorem 8 (Non-Uniqueness of Minimal Interaction Sets for Outcomes)***
*There exist outcomes $o_c \in OC$ and operational contexts $OpsC$ for which multiple distinct minimal interaction sets exist that are each sufficient to determine $o_c$ under $OpsC$.*
**Proof:** Let $o_c$ be an outcome whose determination under $OpsC$ depends on interactions that can occur through alternative admissible interaction structures. By Definition 12, any interaction set sufficient to determine $o_c$ must be a subset of $IR^*$. Suppose there exist two interaction sets $\mathcal{I}_{o_c}^1 \subseteq IR^*$ and $\mathcal{I}_{o_c}^2 \subseteq IR^*$ such that:
1. Each set is sufficient to determine $o_c$ under $OpsC$, and
2. Neither set is a subset of the other.

Such situations arise whenever the operational context admits alternative interaction realizations that independently suffice to establish the same outcome condition (e.g., redundant sensing, alternative actuation paths, or equivalent boundary-crossing exchanges). By Theorem 7, each of $\mathcal{I}_{o_c}^1$ and $\mathcal{I}_{o_c}^2$ admits a minimal subset sufficient for determining $o_c$. Since $\mathcal{I}_{o_c}^1 \neq \mathcal{I}_{o_c}^2$ and neither subsumes the other, the resulting minimal sets are distinct. Hence minimal interaction sets for outcome determination need not be unique.

***Theorem 9 (Non-Essential Interactions May Be Safely Removed)***
Let $\mathcal{O}_d \subseteq OC$ be a set of desired outcomes, and let $\mathcal{C}$ be the set of operational contexts under consideration. Let $(x, y) \in IR$ be an interaction such that, for every $o_c \in \mathcal{O}_d$ and every $OpsC \in \mathcal{C}$, $(x, y)$ does not belong to any minimal interaction set sufficient to determine $o_c$ under $OpsC$. Then removing $(x, y)$ from the problem-space representation does not affect the determinability of any outcome $o_c \in \mathcal{O}_d$ under any operational context in $\mathcal{C}$.
**Proof:** By Theorem 7, for each $o_c \in \mathcal{O}_d$ and $OpsC \in \mathcal{C}$, there exists at least one minimal interaction set $\mathcal{I}_{o_c}^{min} \subseteq IR^*$ sufficient to determine $o_c$. By hypothesis, $(x, y)$ does not appear in any such minimal set for any $o_c$ or $OpsC$. Therefore, for every outcome-context pair, there exists a sufficient interaction set that does not rely on $(x, y)$. Removing $(x, y)$ from the representation preserves at least one sufficient interaction set for determining each desired outcome under each context. Consequently, outcome determinability is preserved. Hence the interaction $(x, y)$ is non-essential with respect to $\mathcal{O}_d$ and $\mathcal{C}$, and may be safely removed without loss of reasoning sufficiency.

***Theorem 10 (Outcome Truth is Boundary Independent; Outcome Classification is Not)***

Let $o_c \in OC$ be an outcome evaluated under operational context $OpsC = (EnvInput, IR^*)$. Let $B_1 = (U_{int}^1, U_{ext}^1)$ and $B_2 = (U_{int}^2, U_{ext}^2)$ be two system boundaries defined over the same problem-space world, corresponding to different choices of system-of-interest.

Then:
1. The truth value of $o_c$ under $OpsC$ is boundary-independent under boundary re-selection: $o_c(OpsC)$ is the same whether evaluated relative to $B_1$ or $B_2$.
2. The classification of $o_c$ as internal or external may differ between $B_1$ and $B_2$. Specifically, if there exists an interaction $(x, y) \in \mathcal{I}_{o_c}$ such that $(x, y)$ is internal relative to $B_1$ but boundary-crossing relative to $B_2$, then $o_c$ is classified as internal relative to $B_1$ and external relative to $B_2$.

**Proof:** By Definition 12, the truth value of $o_c$ under $OpsC$ is determined by its associated interaction set $\mathcal{I}_{o_c} \subseteq IR^*$ and the flows carried on those active interactions. The active interaction set $IR^*$ is determined by the operational context, not by boundary choice. Therefore, changing the boundary from $B_1$ to $B_2$ does not alter which interactions are active or what flows they carry. Hence $o_c(OpsC)$ remains unchanged. By Definitions 12.a and 12.b, outcome classification depends on whether interactions in $\mathcal{I}_{o_c}$ cross the declared boundary. Boundary re-selection changes which interactions are classified as internal, inbound, outbound, or external. If an interaction $(x, y) \in \mathcal{I}_{o_c}$ has both endpoints (source or destination) inside $B_1$ but one endpoint outside $B_2$, then $(x, y)$ is internal relative to $B_1$ and boundary-crossing relative to $B_2$. Therefore, boundary re-selection may reclassify $o_c$ from internal to external (or vice versa) while preserving its truth value.

### Definition 13 (Stakeholder):
*A stakeholder, denoted $Sh$, is a system (called also an actor) that has a vested interest in the $SysSol$ being developed, or the project being undertaken, or is affected by them. Each stakeholder is associated with a non-empty set of goals, desires, or objectives which attainment is related to the outcomes generated by the $OpsSol$. Formally, for each stakeholder $S$, there exists a set of goals $G = \{g_1, g_3, \ldots, g_i\}$, where $i \geq 1$. Each $g_i \in G$ represents a distinct goal, desire, or objective of the stakeholder.*

### Definition 14 (Desired Outcomes):
*A Desired Outcome, is an external outcome (Definition 12.b) whose satisfaction supports the achievement of at least one stakeholder goal. Let $OC$ denote the set of outcomes (Definition 12), and let $G$ denote the set of stakeholder goals. The set of desired outcomes is denoted: $O_d \subseteq OC$. For each desired outcome $o_d \in O_d$, there exists at least one stakeholder goal $g \in G$ such that: $o_d \Rightarrow g$, where the implication $o_d \Rightarrow g$ denotes semantic sufficiency, meaning that for any Operational Context $OpsC$, $o_d(OpsC) = TRUE \Rightarrow g(OpsC) = TRUE$.*

**Conditions**
- Every desired outcome semantically supports at least one stakeholder goal. Outcomes that do not support any goal are not desired outcomes.
- $O_d \subseteq OC$.
- Desired outcomes express stakeholder intent regarding what should hold, whereas outcomes describe what does hold under a given Operational Context.

***Note:*** Desired outcomes need not be individually necessary or sufficient to satisfy a goal. In general, stakeholder goals may require sets of desired outcomes to hold jointly. The identification of necessary or sufficient outcome sets is a separate analysis and is not assumed in this definition.

Desired outcomes are the stakeholder-relevant subset of external outcomes. Internal outcomes arise within the system boundary and may support or hinder external outcomes, but they are not evaluated by stakeholders directly. External outcomes represent how the system affects its environment or other external systems, and desired outcomes capture which of these external effects must hold to satisfy stakeholder goals.

*Theorem 11 (Sufficiency in the Semantic Problem Space World)*
*A problem-space representation is sufficient for reasoning if and only if, for every desired outcome and every operational context under consideration, the representation provides the boundary semantics, interaction structure, admissibility conditions, and outcome grounding required to determine the truth value of that outcome under that context.*

Let $O_d \subseteq O$ be the set of desired outcomes. A problem-space representation is sufficient for reasoning about $O_d$ over a set of operational contexts if and only if, for every $o_c \in O_d$ and every operational context $OpsC = (EnvInput, IR^*)$ under consideration, the representation defines:
1. a system boundary $B_S = (|SysSol|, ES \cup Env)$
2. the interaction relation $IR$ and its boundary-relative subsets $IR_{int}, IR_{in}, IR_{out}, IR_{ext}$ (relative to $B_S$)
3. admissibility conditions for each receiving entity $y$ via $Dom(f_y)$
4. the outcome association $I_{o_c} \subseteq IR^*$ used to determine $o_c$ under $OpsC$

Under these conditions, the truth value of $o_c$ under $OpsC$, denoted $o_c(OpsC)$, can be determined from the problem-space representation for all $o_c \in O_d$ and all operational contexts under consideration.

*Proof*:
- Assume the representation is sufficient for reasoning about $O_d$. Then for each $o_c \in O_d$ and each $OpsC$ under consideration, the truth value $o_c(OpsC)$ must be determinable from the representation. By Definition 12 (Outcome), determining the truth value of $o_c$ under $OpsC$ requires an associated interaction set $\mathcal{I}_{o_c} \subseteq IR^*$ and the interaction evidence carried on those active interactions. To interpret 'active,' 'boundary-crossing,' and 'admissible,' the representation must provide the system boundary $B_S$ (Definition 9.a), the interaction relation $IR$ and its boundary-relative classification (Definitions 3.a–3.c), and admissibility via $Dom(f_y)$ (Definition 3). Hence constructs (1)-(4) must be present.
- Assume constructs (1)-(4) are present. Let $o_c \in O_d$ and let $OpsC = (EnvInput, IR^*)$ be any operational context under consideration. By (4), the interaction set $\mathcal{I}_{o_c} \subseteq IR^*$ used to determine $o_c$ under $OpsC$ is defined. By (1)-(2), each interaction in $\mathcal{I}_{o_c}$ is interpretable relative to the system boundary. By (3), any flow used as evidence in $\mathcal{I}_{o_c}$ must satisfy the admissibility constraint at its receiving entity. Therefore, the interaction evidence required by Definition 12 is fully specified within the representation, and $o_c(OpsC)$ can be determined. Since $o_c$ and $OpsC$ were arbitrary, the same holds for all $o_c \in O_d$ and all operational contexts under consideration.

*Corollary 1 (New Desired Outcomes May Break Prior Sufficiency Claims on the Problem Space)*
*Sufficiency of a problem-space representation with respect to a set of desired outcomes need not be preserved when new desired outcomes are introduced.*

Let $O_d \subseteq O$ be a set of desired outcomes for which a problem-space representation is sufficient. If the desired outcome set expands from $O_d$ to $O_d \cup \{o_{new}\}$, a representation that was sufficient for reasoning about $O_d$ need not be sufficient for reasoning about $O_d \cup \{o_{new}\}$.

*Proof:* By Theorem 11, a problem-space representation is sufficient for reasoning about a desired outcome set if and only if the sufficiency conditions (1)-(4) hold for every outcome in that set under the operational contexts under consideration. Even if these conditions hold for all $o_c \in O_d$, they may fail to hold for the newly introduced outcome $o_{new}$. In particular, the representation may not specify the outcome association $\mathcal{I}_{o_{new}} \subseteq IR^*$, or may lack the boundary semantics or admissibility information required to determine $o_{new}$ under the relevant operational contexts. Therefore, sufficiency for $O_d$ does not imply sufficiency for $O_d \cup \{o_{new}\}$.

*Definition 15 (Functional Requirement):*
*A functional requirement is a statement θ that prescribes the required transformation of an input to an output by the*

*SysSol (adopted from [9]).*

$$\theta \subseteq I \times O$$

**Conditions:**
- *Requirements are only concerned with the SysSol, and not the external systems and their interactions.*
- *Requirement statements can be recursive to enable further decomposition.*

***Note:*** Requirements are prescriptive in nature while functions are descriptive that realize the prescribed transformation. Without getting any formal semantics, a requirement $\theta$ can be expressed as follows:

$\theta$ = *The <SS1> shall <I/O transformation> under <conditions>.*

### Definition 16 (Functions):
*A function is an abstract transformation that maps admissible inputs to outputs. Formally, a function $f: Dom(f) \rightarrow Cod(f)$ specifies how an entity in the System Solution transforms each admissible input into an output. If the entity maintains internal state, its state evolution under inputs is captured by the associated state-transition mapping $\delta_f: St(f) \times Dom(f) \rightarrow St(f)$.*

**Conditions:**
1. *Every system element in SysSol performs at least one function.*
2. $|Dom(f)| \geq 1, |Cod(f)| \geq 1.$ *Each function must accept at least one admissible input and produce at least one output.*
3. *For every admissible input $i \in Dom(f)$, the output $f(i)$ belongs to the codomain: $f(i) \in Cod(f)$. This ensures the system's input–output mappings remain within the defined domain and codomain.*

***Note:*** Although a function defines an abstract input–output transformation, its realized effects in an operational setting depend on the interactions defined in $OpsSol = (SysSol, ES, IR)$.

Let us consider the following example traffic light system abstracted by a function $F$. This example is to better understand all the fundamental elements discussed above. To keep it simple, an exhaustive list of problem space constructs are not discussed, but a few relevant ones in each set so that readers can see how the various definitions (Definitions 1 to 16) fit together in practice. The objective is to provide a concrete scenario that clarifies how these definitions interrelate, giving readers a tangible sense of the theory's applicability and to achieve a shared alignment of mental models.

### Example 1 (Traffic Control System)

*Table 1. Traffic Control System Elements (Part 1)*

| | | |
|---|---|---|
| Stakeholder ($Sh$) | $Sh_1$: City Traffic Department | $Sh_2$: Pedestrians |
| Goals ($G$) | $g_{11}$: need to minimize accidents | $g_{21}$: need to ensure crossing safety |
| | $g_{12}$: need to reduce congestion | |
| Desired Outcome ($O_d$) | $o_d$: Safe traffic flow and rules maintained | |
| Functional Requirement ($\theta$) | $\theta_1$: The system shall transform timer input into appropriate signal output (Red, Yellow, Green) under all defined operational conditions (Peak, Night) to maintain safe traffic flow. | |

In this traffic control system example, two operational contexts ($OpsC$) are identified along with two external systems, vehicles, and pedestrians. These external systems along with the environment are outside the system boundary ($B_s$). The clock is identified as the subsystem of the traffic light ($SysSol$) and is considered inside the system boundary ($B_s$). $OpsC_1$ ($T_{Day}, IR_1^*$) represents the system operating during the day under high traffic volume, with relatively short signal cycles to maximize throughput at rush hours during the day. $OpsC_2$ ($T_{Night}, IR_2^*$) applies during the night when traffic density is low, with longer green phases, aiming to conserve energy while maintaining basic safety overnight. $\{T_{Day}, T_{Night}\}$ are considered as the inputs from the environment that flows into the system boundary and activates a set of distinct interactions.

Table 2. Traffic Control System Elements for $OpsC_1$ (Part 2)

| | SysSol: Traffic Control System (performs function) | $E_1$: Vehicles | $E_2$: Pedestrians | Sub_SysSol: Clock |
|---|---|---|---|---|
| **Inputs (I)** | $i_1$: Timer trigger | $i_2$: Signal color | $i_{31}$: Signal color | $i_4$: Current time |
| | | | $i_{32}$: Car speed | |
| **Outputs (O)** | $o_1$: Signal color | $o_2$: Car speed | $o_3$: Position | $o_4$: Timer trigger |
| **States (S)** | $s_{11}$: Red | $s_{21}$: Moving | $s_{31}$: Wait | $s_{41}$: Peak |
| | $s_{12}$: Yellow | $s_{22}$: Stop | $s_{32}$: Walk | $s_{42}$: Night |
| | $s_{13}$: Green | | | |
| **State transitions ($\tau$)** | $\tau_1(s_{11}, i_1)$ = Green | $\tau_2(s_{21}, i_2)$ = Stop | $\tau_3(s_{31}, (i_{31}, i_{32}))$ = Walk | $\tau_4(s_{41}, i_4)$ = Night |
| | $\tau_1(s_{13}, i_1)$ = Yellow | $\tau_2(s_{22}, i_2)$ = Moving | $\tau_3(s_{32}, (i_{31}, i_{32}))$ = Wait | $\tau_4(s_{42}, i_4)$ = Peak |
| | $\tau_1(s_{12}, i_1)$ = Red | | | |
| **OpsC = (EnvInput, OpsSol)** | $EnvInput = \{T_{Day}\}$ <br> $OpsSol = (F_1, \{car, pedestrian\}, IR_1^*)$ | | | |
| **Outcome (OC)** | $oc_{1,1}$: signal changes color | $oc_{1,2}$: $E_1$ stops at red | $oc_{1,3}$: $E_2$ cross at red | $oc_{1,4}$: time continues to update |
| | $oc_{2,1}$: shorter green during $OpsCtx_1$ | $oc_{2,2}$: $E_1$ starts at green | $oc_{2,3}$: $E_2$ wait at green | |
| | $oc_{3,1}$: longer green during $OpsCtx_2$ | | | |
| **Note:** $oc_{1,4}$ is an internal outcome generated within $B_s$, while the rest are the observed external outcomes outside $B_s$ | | | | |
| **High-level Outcome ($OC_H$)** | $oc_H$: Safe traffic flow and rules maintained. | | | |
| **Note:** For any operational context, the set of outcomes OC established by the realized interactions $IR^*$ may be abstracted into a high-level outcome set $OC_H$ (i.e., $OC \rightarrow OC_H$) that captures the net effect of those outcomes at the level of stakeholder interpretation. If $OC_H = O_d$, the SysSol meets the stakeholder goals/needs and is thus validated. | | | | |

Section IV established the problem space as a semantic world model by introducing the formal constructs needed to reason about the domain prior to any solution commitment. In particular, the definitions make the system boundary a semantic commitment, treat interactions as admissible flow-bearing relations rather than assumed connectivity, and define outcomes as context-grounded propositions whose truth values are determined by realized interaction patterns. Together, these constructs separate what is true of the operational world from what is later chosen as a solution, and they provide a disciplined basis for attributing consequences to explicit boundary-crossing and internal interactions rather than to implicit assumptions or narrative interpretation. With this semantic core in place, the next section examines the significance of the developed theory

## IV. SIGNIFICANCE OF THE THEORITICAL FRAMEWORK

This section introduces the significance of the developed problem space theory from a practitioner's perspective. In this paper, the term '*problem space*' is not used as an informal stereotype for '*the set of needs*' or '*the stakeholder goals*', but as a formally interpretable semantic world model. It is a domain in which constructs like boundaries, entities, interactions, operational context, and outcomes have explicit meaning and can support disciplined reasoning. The term '*semantic world model*' emphasizes that the problem space is not treated as an informal collection of shared beliefs, heuristics, or common knowledge. Instead, it is made explicit as a formally defined domain whose core constructs are assigned traceability and accountability so that claims about what follows under a given operational context are derived from the model itself. Such a rigorous formal world avoids implicit assumptions, inherited solution bias, or dependency on prescriptive artifacts that prematurely embed design commitments. In this framing, disciplined reasoning occurs prior to solution design by separating what is true of the problem domain (boundaries, external entities, admissible interactions, contextual conditions, desired outcomes) from what is later chosen as a solution (physical architectures, implementations, and system requirements).

While the paper establishes a full set of formal definitions for constructing this world model, the discussion here is intentionally centered on the three constructs that constitute its semantic core: interactions (Definition 3), boundaries (Definition 9) and outcomes (Definition 12). The remaining definitions are not treated as independent focal points in this section. They are introduced as supporting constructs that enrich and constrain these three core notions by making admissibility explicit, enabling context-dependent activation, and grounding outcome evaluation in the world model.

*Definition 3 (Interactions):*
Two reasoning failures recur in early problem formulation. First, engineers assume that connection implies flow, i.e., "if A is connected to B, then A will affect B." This conflates structural possibility with operational reality. A sensor may be wired to a controller, but whether a signal propagates depends on whether the controller can interpret it. Second, engineers discuss interactions without stating their boundary-relative classification: Is this interaction internal to the system? Entering from outside? Leaving to an external entity? The same interaction is often treated as "internal" when justifying design choices but "external" when deflecting accountability. Definition 3 addresses both failures. It requires every interaction to have an explicit admissibility condition: a flow propagates along $(x, y)$ only if the receiving entity $y$ has a function $f_y$ whose domain includes that flow. Connection is no longer sufficient; the receiving side must be capable of interpreting what is sent. The definition further requires every interaction to be classified relative to the system boundary (Definitions 3.a–3.c): internal, inbound, outbound, or external. This classification is fixed once the boundary is declared—it cannot shift mid-argument. Admissibility grounds interactions in defined input domains rather than tacit assumptions, making flow claims auditable. Boundary-relative classification fixes accountability at the point of problem formulation, preventing engineers from reclassifying interactions opportunistically to support preferred conclusions. Together, these conditions transform interactions from informal notions of "connection" into formal commitments that support consistent reasoning and defensible traceability.

*Definition 9 (Boundary):*
Boundary ambiguity is among the most common sources of reasoning failure in systems engineering. The same entity is treated as "in scope" when justifying a requirement or claiming credit for an outcome, but "out of scope" when failure attribution, verification evidence, or traceability is demanded. This inconsistency arises because boundaries are treated as informal modeling conveniences, i.e., lines drawn for diagrammatic clarity rather than semantic commitment. The problem is compounded when requirements and architectures are written before the operational context is explicit, and the boundary shifts to accommodate whatever argument is being made at the moment. Definition 9 establishes the boundary as a formal partition of the entity universe. Every entity belongs to exactly one of three classes: the System Solution (| *SysSol* |), External Systems (*ES*), or the Operational Environment (*Env*). There is no overlap and no ambiguity. Once declared, this partition determines which interactions are internal, which cross the boundary, and which are entirely external. Definition 9.a further specifies that the system boundary induces a classification of all interactions relative to the chosen system-of-interest. By requiring explicit, mutually exclusive classification, the definition eliminates scope drift. An entity cannot be "in scope" for credit and "out of scope" for accountability, it is inside or outside, and that classification is fixed at the point where the problem is defined. This transforms the boundary from an informal convenience into a semantic commitment with downstream consequences for interaction classification, outcome attribution, and sufficiency evaluation.

*Definition 12 (Outcomes):*
Engineers routinely state outcomes as if they were unconditional facts: "Patient safety ensured," "congestion is reduced," "mission success is achieved." These statements suffer from two defects. First, they have no truth conditions, i.e., under what operational context is this claim evaluated? "Patient safety" might hold under normal operation but fail under sensor degradation. Without specifying context, the claim is untestable. Second, they have no grounding, i.e., what interactions make this the system's outcome rather than a coincidental world state? An outcome that cannot be traced to system-participating interactions is not an outcome of that system, and the system cannot be held accountable for it. Definition 12 addresses both defects. It treats an outcome as a proposition whose truth value is evaluated only relative to a specified operational context $OpsC$. Asking "is $o_c$ true?" without specifying context is meaningless within the theory. The definition further requires that every outcome be grounded in an explicit interaction set $\mathcal{I}_{o_c} \subseteq IR^*$ that includes at least one interaction where the System Solution participates. This prevents "floating" outcomes, i.e., claims about the world that sound like system effects but lack any causal path from system behavior.

The association between an outcome and its grounding interaction set is a modeling commitment, not a derived causal relationship. The theory does not infer which interactions cause which outcomes; it requires the modeler to declare that association explicitly. This declaration is a substantive claim about the problem domain, one that can be challenged, refined, or validated through domain analysis. The theory provides the representational structure for such claims; domain expertise provides their content.

With this semantic framing established, the rest of the section summarizes the practical significance of the developed axioms, theorems, and corollaries, emphasizing how they collectively resolve common problem concerns. This is done by making what is assumed about the world explicit, traceable, and analyzable before prescriptive requirements and solution design choices are introduced.

*Axiom 1:* This axiom establishes an explicit modeling commitment that all reasoning about causality, boundary-crossing interactions, and outcome attribution is valid only with respect to the declared problem-space world. Phenomena not represented within this boundary are treated as semantically irrelevant for the purposes of analysis, even if they may exist in reality. In practice, engineering reasoning often relies on implicit or unspecified external influences, which undermines traceability and renders conclusions interpretation-dependent. By enforcing a closed-world boundary, the framework ensures that every claim, dependency, and inference is grounded in explicitly modeled constructs within the problem-space world.

*Axiom 2:* Systems engineering reasoning is inherently multilevel and recursive: systems may be decomposed into subsystems or aggregated into higher-level systems depending on the analytical objective. This axiom establishes boundary declaration as a necessary semantic commitment, ensuring that statements about admissible interactions, operational contexts, and outcome attribution are logically well-formed rather than interpretation-dependent.

*Axiom 3:* Outcomes are not intrinsic or universal properties of a system in isolation. Instead, they are propositions whose truth depends on the environmental inputs and the set of interactions that are realized under a given operational context. By anchoring outcome evaluation to explicit operational contexts, this axiom ensures that outcome semantics are well-defined, repeatable, and independent of informal narrative or unstated assumptions.

*Theorem 1:* This theorem formalizes how the operational environment can affect the system behavior. Practitioners routinely reason about environmental effects, but those effects are often treated informally as '*conditions that matter*' without a clear mechanism of influence. The theorem establishes that environmental influence becomes engineering-relevant only when it enters through inbound boundary-crossing interactions as admissible inputs and may then drive state transitions. This resolves a common gap in problem definition as it becomes possible to distinguish environmental inputs, their effects on the operational context and system behavior.

*Theorem 2:* This theorem enforces the distinction between interactions that exist conceptually and interactions that occur in a specific operational context at a particular instance. In practice, system models specify all potential interactions, and problem statements implicitly treat them as always active. The theorem prevents this mistake by making interaction realization explicitly dependent on environmental inputs and operational context. This provides the formal basis for context-based reasoning, i.e., practitioners can reason about which interactions are relevant under specified conditions and that supports the realization of external outcomes, rather than treating the entire set of interactions as operational facts.

*Theorem 3:* This theorem ensures that boundary semantics are not ad hoc conventions that must be reinvented at each level of decomposition. When practitioners shift focus from a system to a subsystem, or aggregate subsystems into a higher-level view, the same formal rules for classifying interactions as internal, inbound, outbound, or external apply. Admissibility constraints carry through unchanged. This supports consistent multi-level reasoning, where an engineer analyzing a subsystem applies the same semantic framework as one analyzing the full system, preserving traceability and preventing the definitional drift that often occurs when different teams work at different levels of abstraction.

*Theorem 4:* This theorem protects practitioners from a common failure in early problem formulation caused by treating architectural connectivity as evidence of operational behavior. The theorem states that activity is context-dependent and that interface availability is not sufficient to conclude occurrence. This is a direct resolution to one of the conceptual gaps mentioned in the introduction section, i.e., the need to separate what can happen from what does happen under specified operational conditions. With this result, claims about outcomes must be anchored to an operational context rather than inferred from structure alone.

*Theorem 5:* This theorem establishes a strict accountability rule for operational reasoning in the problem space world, i.e., only what is realized under the stated operational context can be used to justify claims of outcomes. Practitioners often encounter arguments where a proposed effect is defended by pointing to implicit interaction relations that are not active under the context being analyzed. This theorem avoids that reasoning. It forces '*affect*' to be supported by an active interaction carrying an admissible flow under the context in operation. The practical result is that operational contextual reasoning becomes repeatable and defensible because it is constrained to explicitly realized interactions rather than implicit causal narratives.

*Theorem 6:* This theorem establishes a locality principle for outcome reasoning: if the interactions that matter for an outcome are the same across two contexts, then the outcome's truth value is the same, regardless of what else differs between those contexts. This is powerful for verification planning and context-based reasoning. Practitioners can identify which aspects of operational context are relevant to a given outcome and which are not. If two operational contexts preserve the same outcome-grounding interaction set $\mathcal{I}_{o_c}$, then the verification evidence established for $o_c$ in one context is reusable for the other, since the basis for concluding $o_c$ has not changed. Any additional active interactions in the second context do not break that reuse; they only require additional verification activity for the other outcomes whose truth values are grounded in these additional interactions. As a result, verification effort is focused where the semantics actually change, reducing unnecessary test execution thereby saving time and cost.

*Theorem 7:* This theorem establishes that outcome reasoning does not require exhaustive modeling of all possible interactions. For any outcome and operational context, there exists at least one minimal set of interactions that suffices to determine whether the outcome holds. This enables focused problem-space modeling, supports reduction of unnecessary detail, and prevents over-constraining early analyses with interactions that are irrelevant to outcome determination.

*Theorem 8:* This theorem formalizes a common operational reality: the same stakeholder-relevant outcome can be established through more than one distinct '*evidence path*' in the problem-space world. In practice, engineers routinely introduce redundancy, alternate interaction path, or multiple admissible domains to achieve the same effect. The result is that outcome reasoning should not assume a single strict interaction chain. Instead, verification and traceability must explicitly account for multiple minimal interaction sets that can each justify the same outcome under the same context. This prevents false claims of incompleteness when one interaction grounding is absent, while ensuring that all admissible interaction groundings that can determine the outcome are explicitly recognized, assessed, and managed within the problem formulation. Practitioners may then select the most suitable grounding based on project constraints such as cost, schedule, and operational burden.

*Theorem 9:* This theorem provides a formal basis for model reduction. As problem-space representations grow in complexity, practitioners face the challenge of distinguishing essential structure from incidental detail. This theorem gives a precise criterion: an interaction is non-essential if it does not appear in any minimal set for any desired outcome under any operational context of interest. Such interactions can be safely removed without affecting the ability to reason about desired outcomes. This supports model simplification, reduces verification scope, and helps practitioners focus attention on the interactions that actually matter for stakeholder-relevant outcomes. Conversely, any interaction that appears in at least one minimal set for at least one desired outcome under at least one context is essential and must be retained.

*Theorem 10:* This theorem separates two concerns that are often conflated in practice: outcome truth and outcome classification. Truth is a semantic property determined by active interactions and flows under an operational context; it does not depend on where the practitioner draws the system boundary. Classification is a structural property that determines accountability, verification scope, and interface requirements, it depends entirely on boundary choice. When practitioners shift focus from a system to a subsystem (or vice versa), they are re-selecting the boundary. This theorem assures them that outcome truth is preserved under such shifts; only the classification changes. An outcome that was "internal" at the system level becomes "external" when reasoning about a subsystem, but what is true about that outcome remains unchanged. This enables consistent multi-level reasoning without the risk of inadvertently changing what the model says is true.

*Theorem 11:* This theorem provides a formal criterion for representational sufficiency, a question that practitioners typically answer by intuition or precedent. Rather than asking "*do we have enough detail*?" in the abstract, this theorem specifies exactly what "enough" means: the representation must support truth-value determination for every desired outcome under every relevant operational context. This gives engineers a concrete checklist: *Have we defined the boundary? The interactions? The admissibility conditions? The outcome-interaction associations?* If any of these are missing for a desired outcome under any context of interest, the representation is insufficient and reasoning cannot proceed

*Corollary 1:* This corollary formalizes a central reality of practice: problem-space world model reasoning is iterative. New stakeholder concerns routinely expand the set of desired outcomes and, in doing so, can invalidate earlier sufficiency claims. Rather than treating this as an engineering failure, the corollary makes it a predictable consequence of sufficiency being outcome-relative. This implies, when the desired outcome set expands, the representation may require revised boundary commitments, additional interactions, updated admissibility conditions, and new outcome grounding relationships. The practical impact is that iteration is no longer performed by informally modifying problem-space constructs while leaving the rationale implicit. Instead, the corollary makes outcome changes actionable: when the desired outcome set is modified, it specifies exactly what must be updated in the problem-space model and how that change propagates through boundaries, interactions, admissibility conditions. This replaces informal heuristics or prior beliefs that the prior formulation still applies, and instead provides a disciplined basis for determining which constructs are impacted and why.

Table 3 concludes this section by summarizing how each research question is formally answered by the developed axioms, definitions, theorems, and corollaries, establishing explicit traceability from the problem-space semantic constructs to the paper's stated research objectives.

*Table 3: Relation between developed theoretical constructs and proposed research questions*

| RQ | Axioms/Definitions/Theorems/Corollary | Description |
|---|---|---|
| RQ1 | *Axioms*: A1, A2<br>*Definitions:* 3, 3.a-3.c, 4-6, 9, 9.a, 12a-12.b<br>*Theorem:* 3, 10 | Boundaries are semantic commitments: every entity is inside or outside, every interaction has a unique boundary-relative class, and re-selecting the boundary may change classification but not truth. |
| RQ2 | *Axioms:* A1, A3<br>*Definitions:* 3, 6-8, 10-12<br>*Theorems:* 1, 2, 4, 5, 6, 7, 8 | Operational context selects the active interaction set $IR^*$, and outcomes are grounded only in explicit interaction sets sufficient under that context, blocking structure- or narrative-based reasoning. |
| RQ3 | *Axioms:* A1, A2, A3<br>*Definitions:* 9.a 11, 12, 14<br>*Theorem:* 6, 9, 10<br>*Corollary:* 1 | Sufficiency is outcome- and context-relative: it holds exactly when desired outcomes are determinable across contexts, can break when outcomes expand, and is preserved under safe reduction and boundary re-selection. |

## V. THEORY-GUIDED PROBLEM SPACE REASONING IN PRACTICE

The following section provides a reasoning-oriented discussion of how the proposed formal problem space semantics can be used in practice for a hypothetical system *A*. Rather than presenting a traditional case study tied to a specific domain, we structure the example as a dialogue between a stakeholder and an engineer who are attempting to frame a problem correctly before any solution commitments are made. This is intentionally in a non-traditional case study format: the goal is to 'tell a story about identifying the problem itself rather than proposing a solution,'. It shows how the theory prompts practitioners to ask the right questions while reasoning about the problem space world.

As the practitioners work through defining the problem, the definitions and theorems introduced in Sections III convert implicit notions assumed at the early system life cycle into formal commitments that must be stated and checked within the world model. In doing so, the theory functions as a disciplined reasoning criterion by forcing assumptions to become explicit, replacing narrative justification with interaction-based evidence, and providing a principled basis for judging whether the current problem space representation is sufficient for reasoning about desired outcomes across all operational contexts. The following dialogue is therefore written to make these explicit.

*Stakeholder:* "We need System *A (Definition 4)* to meet Goal *G (Definition 13)* under certain operational conditions. For example, when scenario *X* occurs in the operational world, we expect the system to behave in a way that supports that goal. *Can the new theory be used to make sure the problem is framed correctly?*"

*Engineer:* "Absolutely. We first begin by translating the goals to a list of desired outcomes (*Definition 14*) rather than initiating any prescriptive artifact. These are explicit conditions in the operational world whose truth is what will count as '*goal achieved*' in a given context. Before we say anything about what System *A* 'should do,' we ask: *what observable truth must hold to say Goal G is satisfied?* That truth value is formally represented as the desired outcome. Once the desired outcome is established, the next question is *where must this truth hold*? This immediately forces the abstract scenario *X* to be stated as a formal construct named operational context (*Definition 11*), i.e., the environmental inputs and conditions under which we will observe the outcomes (*Definition 12*).

*Note: The outcomes realized under the defined relevant operational contexts are later evaluated against these set of desired outcomes (Definition 14) to determine whether the stakeholder goals are satisfied.*

*Stakeholder:* "System *A* should achieve Outcome *Y*. *Can we not write that as a prescriptive artifact (requirement)?*"

*Engineer:* "Not yet. In this theory, an outcome (*Definition 12*) is a proposition whose truth is evaluated *only* relative to an operational context. If we write '*A* shall achieve *Y*' without stating the context, an implicit claim '*Y* holds in general', is being made. That is exactly what the outcome semantics forbids because it collapses context-dependent truth into an unconditional statement."

*Stakeholder:* "So, goal G is the driver, and success is checked against desired outcomes. If we cannot state the desired outcome as an explicit condition with a truth value in the operational world, then we have not yet specified what '*goal achieved*' means. The abstract vague scenario *X* is also formalized as an operational context involved with the system-of-interest (*Definition 4*)"

*Engineer:* "Correct, but it should be noted that while we formalize the context, we must parallelly ask the question: *what, precisely, is the system-of-interest whose behavior we will hold accountable for realizing the outcomes in the operational context?* That question is not well-posed without first committing to a system boundary (*Definition 9.a*)."

*Stakeholder:* "Meaning we must decide what we are calling '*System A*' versus what we are treating as external (*Definition 5*) or environmental (*Definition 6*), otherwise we will not know what interactions (*Definition 3*) are allowed to be attributed to the system."

*Engineer:* "Yes. If the boundary is not fixed, we can keep moving responsibility across the line, calling an entity 'in scope' when it helps and 'out of scope' when it is convenient. The theory prevents that by forcing an explicit boundary commitment before we reason about interactions or outcomes. With this in mind, we are ready to make the system boundary (*Definition 9*) explicit and proceed consistently. This is crucial as the system boundary is a formal semantic commitment that determines scope. *For instance, is there an external System B or environment element that provide an input to A? If so, that entity lies outside A's system boundary.* Everything inside the boundary will be part of System *A* (the solution we are designing), and everything outside represents either external systems or the operational environment."

*Stakeholder:* "Understood. In our problem space world, System *A* includes the internal functions (*Definition* 16) we will design. There is also an External System *B* (*Definition 5*) that exchanges some admissible signals or exchange of material or energy (*Definition 1, 2, 3*) with it, and there are environmental phenomena (*Definition 6*) that can stimulate the scenario *X*. *Since B is connected to A, the system will receive and accept flows from B right?"*

*Engineer:* "Connection alone is not enough. With the boundary set, we can classify interactions (*Definition 3*) unambiguously. Any interaction between *A*'s elements stay internal and any interaction relative to *A* with *B* or Environment crosses the boundary (*Definition 3.b*). We should specify what flows across that interaction, and ensure it is an admissible input to *A*. In other words, the functions in *A* or *B* must be defined to accept that kind of input. The theory prompts us to ask this: "*Have we defined what inputs A can legitimately receive from B?*" If not, we cannot assume any arbitrary signal from *B* will flow and affect *A*. Only a flow that lies in *A*'s acceptable domain will propagate along the interaction (*Definition 1, 2*). Previously, we might have overlooked this, implicitly assuming "if *B* is connected to *A*, it will just work." Now we make it explicit: for example, a *B → A* interaction is only valid when carrying an admissible flow value. This prevents ambiguous or undefined interactions by design."

*Stakeholder:* "*If it is common knowledge or if everyone knows that a particular interaction will affect A, should we spend time and effort formally representing it?*"

*Engineer:* "Absolutely. If it affects *A* in the problem-space world, then it must be represented through an explicit interaction carrying an admissible flow. If left 'known but unmodeled,' we have created a hidden influence channel that can neither be traced nor tested. The theory forces the choice: either represent it, or do not use it to justify any claim."

*Stakeholder:* "Understood. So far, we have identified System *B* feeding inputs into System *A*. *Are there any outputs going back?*"

*Engineer:* "Possibly. If System *A* needs to send something out to influence *B* or the environment, those would be outbound interactions crossing the boundary (*Definition 3.b*). Let us consider an inbound interaction where *B* sends input into *A*, and perhaps an outbound interaction where *A* responds with output back to *B*. Each of these is defined in our interaction set, with specified directions and admissible flow. The formal theory tells us that with the boundary fixed, every interaction in our model falls into a clear category, internal, inbound, outbound, or external, with no overlap (*Theorem 3*). This was often abstracted and missed before; now there is no confusion about what is inside versus outside or which interactions cross the boundary. We have eliminated a common ambiguity: for example, treating an entity as "inside" in one discussion and "outside" in another, relative to a fixed boundary partition, because the system boundary definition (*Definition 9*) forbids that inconsistency."

*Stakeholder:* "That makes sense. Now, about the outcome *Y* (*Definition 12.b*). We want *Y* to happen in a particular scenario. *How do we use the theory to reason about it?*"

*Engineer:* "We will use the conditions given under the construct's, operational context (*Definition 11*) and outcomes (*Definition 12*). An operational context is basically a specific scenario, here, '*X happens*', that provides certain inputs to System *A*. Under that context, only some of the potential interactions will actually become active. The theory forces us to anchor *Y* to explicit scenario conditions rather than heuristics or assumptions. In summary, we identify the active interaction set for the scenario. Other interactions we defined (maybe other inputs or outputs that exist conceptually) remain inactive because they are not triggered by this context, and that is fine. We no longer confuse structural possibilities with actual occurrences (*Theorem 4, 5*)."

*Stakeholder:* "What do you mean by active interaction? We have already defined the $B \rightarrow A$ interaction and reasoned about admissible flows, so it will occur always, right?"

*Engineer:* This is one of the key insights from the theory. Defining a possible interaction (like $B \rightarrow A$) does not mean it is always in use. It becomes active only when a flow from the environment is crossed into the system boundary under a given context. This admissible flow triggers the relevant set of all active interaction (*Theorem 2*). System *A* will receive it, and perhaps change its internal state *(Definition 7)* inducing a state transition *(Definition 8)* in response (*Theorem 1*). Now, with the input processed, *A* might produce some output. If Outcome *Y* is to happen, presumably *A* must output something (or cause some observable effect) that leads to *Y*. That output would travel via an outbound interaction ($A \rightarrow B$ or $A \rightarrow Env$). Crucially, if the context *X* does not occur, say the environmental stimulus never flows, then that inbound interaction stays inactive and *A* might never produce that specific output. Any implicit assumptions on how Y "follows" from A's design is avoided and explicit insights on "*what exact conditions leads to Y*" is formally achieved.

*Stakeholder:* "Right, so we are effectively mapping a cause-and-effect path: in Context *X*, a set of active interactions leads to *Y*. *Now, how do we represent and reason about outcome Y using the theory?*"

*Engineer:* "The theory defines an Outcome (*Definition 12*) as a proposition with an associated truth about a condition that arises in an operational context as a consequence of the interactions. So, outcome *Y* needs to be stated in a verifiable way, something like 'the system interactions with the external entities realizes condition *Y*.' If *Y* is a stakeholder-visible effect, it is an external outcome (*Definition 12.b*), meaning it involves at least one interaction across the system boundary. Otherwise, *Y* is an internal outcome (*Definition 12.a*). The truth of *Y* (did it happen or not) can be evaluated against the set of active interactions realized in the context. The formalism prompts us to ask: "*Which interactions must be active for Y to be true?*" Before, we might have just said '*Y* happens' without evidence; now we trace *Y* to specific model elements (*Theorem 6, 7, 8, 9, 10*)."

*Stakeholder:* "Can we include any stakeholder-visible condition as an outcome even if System A is not attributed with it?"

*Engineer:* "Absolutely not. External outcomes must be grounded in at least one outbound boundary-crossing interaction from the system-of-interest under the context. We cannot have an "outcome" floating around that *A* never influences. This would be a false claim attributed to the system and is treated as outside the closed world boundary (*violates Axiom 1*). In practice, this means if we had some external effect we care about, we ensure *A* has a role in causing or preventing it via interactions. It brings accountability: *Y* is not implicit or untraceable, it comes from *A* being operated in Context *X*. We have removed ambiguity about how Y comes about."

*Stakeholder:* "This is very insightful. I see now that a prescriptive artifact like 'System shall achieve *Y* when *X*,' holds many implicit assumptions or hidden knowledge. But, with the help of the theory, we transform these implicit notions to explicit formal constructs. It feels more complicated, but I can tell it makes our reasoning much more concrete and rigorous. *Is there anything else we can reason about?"*

*Engineer:* "The final step is to consider sufficiency, meaning have we modeled enough of the problem space to reason confidently about *Y* (and any other outcomes)? According to the theory, a problem-space representation is sufficient for reasoning if and only if for every desired outcome and operational context under consideration, the representation provides all the sufficient artifacts needed to determine the truth value of the outcome (*Theorem 11*). Let us apply that. *For Outcome Y in Context X, do we have everything?* We got a defined boundary (so we know what lies in/out and where interactions occur), a set of interactions with admissible flows (so we know how flows travel), an operational context (*X* provides specific input), and an outcome proposition (*Y*) grounded in those interactions. Therefore, the formal semantic world model help reason whether *Y* happens or not in this context. If any of these constructs were missing, we will have a gap."

*Stakeholder:* "Great, so, relative to the realized desired outcomes and the admitted operational contexts one can say the problem space world model is sufficient. *But is that sufficiency constant, or does it change?"*

*Engineer:* "It is crucial to note that sufficiency of the problem space is not constant. We should think about evolution: *what if you introduce a new goal or context tomorrow?* Say you add Outcome *Z* (a new stakeholder desire) or decide to expand the context scope of System *A*. Initially, our model might not account for this new outcome. *Z* may depend on an external entity we did not consider or on a context we did not model. In that case, our current problem space world would not be sufficient for *Z*. The theory explicitly proves that adding a new desired outcome can break a previously sufficient model (*Corollary 1*). We need to update the boundary (maybe include another entity or define a new interaction), or add the additional constructs (*Theorem 11*), to re-establish sufficiency. This provides a rigorous way to handle changing goals instead of iterating the problem space representation in an ad-hoc way. In short, the theory not only helps set up the initial problem space but also guides how to adapt it without losing clarity."

*Stakeholder:* "Now I see the kind of questions I should be asking. This theory eliminates the uncertainty that typically arises from unstated scope, ambiguous representations and implicit assumptions. This gives confidence on determining what is true of the problem domain from what is later chosen as a solution."

## VI. CONCLUSION

The developed theory in this paper provides a formal representation of the problem space world to enable consistent, traceable and sound reasoning. It treats the problem space as a semantic world model: a formally defined domain relative to stakeholder goals, so that claims about "what follows" under a context are derived from the model rather than from implicit assumptions or premature design commitments.

Within this framing, the developed theory establishes a minimal semantic core (interactions, boundaries, outcomes) and the supporting constructs as formal definitions adopted from systems theory [23, 39, 40] augmented with set theory. Propositional logic is used to ensure that these constructs are well-posed for disciplined reasoning. The resulting derived axioms and theorems provide a rigorous basis for (i) treating boundary as a semantic commitment and not an informal convenience, (ii) classifying interactions relative to a chosen system-of-interest boundary, (iii) distinguishing context-realized active interactions from set of all available interactions, (iv) grounding outcome truth in explicit interaction sets rather than narrative interpretation, and (v) characterizing when sufficiency claims hold true or evolves as an outcome and context relative property of a problem-space representation.

Finally, the theory was positioned as a reasoning criterion for practitioners. Inferences show that the theory successfully transforms what is typically implicit in existing problem space formulation into explicit commitments that can be stated, verified, and revised prior to solution design, thereby separating what is true of the problem domain from what is later chosen as a solution.

## A. LIMITATIONS

First, this work is intentionally foundational: it establishes formal semantics for problem-space representation, but it does not yet integrate quantitative models (e.g., uncertainty, time-dependent behavior or performance margins) required to compute function capability, affordability or performance metrics. Second, outcomes are grounded through interaction evidence as defined in the model; therefore, the quality of reasoning remains dependent on the completeness and correctness of the modeled entities, admissibility conditions, and context specifications. This paper does not discuss validity claims in detail. Third, the theory is presented at the level of formal constructs and proofs and is not yet operationalized as a tool-integrated framework (e.g., automated checks inside MBSE environments or automated reasoning using an ontology).

## B. FUTURE WORK

Future work will extend the theory with additional problem-space constructs supporting capability measures associated with the functions. Capability constructs will extend the theory with an explicit bridge to the solution space, providing a formal basis for translating stakeholder needs and performance requirements into measurable constraints on the semantic problem space world. Second, the framework will include explicit operations for composition and decomposition of functions so that the same semantics can be applied consistently across multiple levels of abstraction or nested system boundaries. Next, the problem space verification and validation will be formally defined and incorporated into the semantic problem space world. Finally, we will focus on operationalizing the semantic problem space world model as an executable and tool-supported framework that enable automated reasoning. This includes (i) developing an ontology and reasoning layer that can automatically enforce boundary commitments, admissibility constraints, interaction classification, and outcome grounding; and (ii) integrating these checks with MBSE artifacts so that problem space consistency can be assessed prior to requirements and architecture development

## ACKNOWLEDGEMENTS

The author acknowledges the use of ChatGPT to check grammar, spelling, generate figures and paraphrase some text in the paper.

This paper is based upon work supported by the National Science Foundation under CAREER Award No. 2441775 Any opinions, findings, and conclusions or recommendations expressed in this material are those of the authors and do not necessarily reflect the views of the National Science Foundation.